\documentclass[aps,twocolumn,showpacs]{revtex4-1}
\usepackage{amssymb}
\usepackage{amsmath}
\usepackage{amscd}
\usepackage{graphicx}
\usepackage{subfigure}
\usepackage{float}
\usepackage{array}
\newcommand{\PreserveBackslash}[1]{\let\temp=\\#1\let\\=\temp}
\newcolumntype{C}[1]{>{\PreserveBackslash\centering}p{#1}}
\newcolumntype{R}[1]{>{\PreserveBackslash\raggedleft}p{#1}}
\newcolumntype{L}[1]{>{\PreserveBackslash\raggedright}p{#1}}
\begin{document}

\title[Short Title]{Cooling distant atoms into entangled state via coupled cavities}

\author{Li-Tuo Shen$^{1}$}
\author{Xin-Yu Chen$^{1}$}
\author{Zhen-Biao Yang$^{2}$}
\author{Huai-Zhi Wu$^{1}$}
\author{Shi-Biao Zheng$^{1}$}
\email{sbzheng11@163.com}

\affiliation{$^{1}$Lab of Quantum Optics, Department of Physics,
Fuzhou University, Fuzhou 350002, China\\$^{2}$Key Laboratory of
Quantum Information, University of Science and Technology of China,
Chinese Academy of Sciences, Hefei 230026, China}

\begin{abstract}
We propose a scheme for generating steady entanglement between two
distant atomic qubits in the coupled-cavity system via laser
cooling. With suitable choice of the laser frequencies, the target
entangled state is the only ground state that is not excited by the
lasers due to large detunings. The laser excitations of other ground
states, together with dissipative processes, drive the system to the
target state which is the unique steady state of the system.
Numerical simulation shows that the maximally entangled state with
high fidelity can be produced with presently available
cooperativity.
\end{abstract}

\maketitle

\noindent

The major difficulty in the implementation of a quantum processor is
the decoherence due to coupling to the environment. Especially for
the entanglement between distant nodes, the coherence usually
becomes very fragile when it needs to be manipulated externally.
This problem can not be fully solved just based on unitary dynamics
\cite{PRA2008-78-063805,LPR2008-2-527,PRL-2000-85-2392,PRA-1999-59-2468,PRL-2003-90-097902}.
Recently, the dissipation has been used as a resource to generate
the long-time entanglement
\cite{NJP-2009-11-083008,PRA2011-84-022316,PRL2011-106-090502,NJP-2012-14-053022,PRA-2012-85-032111,
JOSAB2011-28-228,PRL2011-107-120502,EPL-85-20007,PRL2011-106-020504,arXiv:1005.2114,PRA2012-85-042320,PRA2011-84-064302}.
One recent experiment has demonstrated the dissipative preparation
of entangled steady-state \cite{PRL2011-107-080503}. Approaches
based on this idea do not require photon detection, observation of
macroscopic fluorescence signals, and definite control of initial
state and evolution time, which are very robust against moderate
environment noise. Vacanti and Beige \cite{NJP-2009-11-083008}
showed the possibility of preparing highly entangled states by
methods being close analogy to laser sideband cooling. Busch
\emph{et al.} \cite{PRA2011-84-022316} proposed an improved
entanglement cooling scheme for two atoms inside an optical cavity
with a rather low cooperative parameter. We have proposed a scheme
for producing entangled states between two atoms trapped in two
coupled cavities in the steady state \cite{PRA2011-84-064302}. The
scheme is based on suppression of the effective decay of the target
entangled state induced by cavity loss by suitably setting the
cavity detuning.

In this paper, we propose an alternative scheme to prepare the
highly entangled steady-state for two atoms trapped in two coupled
cavities based on the idea of laser cooling. Unlike the scheme of
Ref. \cite{PRA2011-84-064302}, the target entangled state becomes
the steady state due to the suppression of its laser excitation.
Through suitable choice of the laser frequencies, the target
entangled state is not affected by the classical drivings due to
large detunings, while each of the other ground states are
resonantly coupled to the corresponding dressed excited states of
the whole atom-cavity system by one laser field. Each excited state
decays to the target state and other ground states due to
dissipative dynamics. As a result, the target state is the unique
steady state. We find a distinct improvement in the fidelity
compared with the previous scheme
\cite{PRA2011-84-064302,PRL-2003-91-177901}. The distributed
entangled steady-state with the fidelity above $90\%$ can be
obtained after a moderate evolution time even when the cooperative
parameter is only $50$, which is impossible by previous methods
\cite{PRA2011-84-064302,PRL-2003-91-177901}. This present work may
constitute an important step toward the realization of quantum
networks with current experimental technology for the coupled-cavity
system \cite{NPHOTONICS2011-6-56,Nature2007-445-896}.

The experimental setup consists of two identical $\Lambda$-type
atoms trapped in two directly coupled cavities respectively, as
shown in Fig. 1. The level $|i\rangle$ of each atom has the
corresponding energy $w_{i}$ ($i=0,1,2$). Without optical laser
driving, the Hamiltonian of the whole system can be written as:
\begin{eqnarray}\label{e1}
H_{NL}&=&\sum_{j=1}^{2}\sum_{i=1}^{2}w_{i}|i\rangle_{jj}\langle i|
+\sum_{j=1}^{2}w_{a}a^{\dagger}_{j}a_{j}+J(a_{1}^{\dagger}a_{2}\cr&&+a_{1}a_{2}^{\dagger})
+\sum_{j=1}^{2}g\big( |2\rangle_{jj}\langle
1|a_{j}+|1\rangle_{jj}\langle 2|a_{j}^{\dagger}\big),
\end{eqnarray}
where $a_{j}$ and $a_{j}^{\dagger}$ are the annihilation and
creation operators for the $j$th cavity field mode with frequency
$w_{a}$ respectively. We here have set the energy $w_{0}$ of level
$|0\rangle$ to be zero. The $j$th cavity field mode resonantly
couples to the $|1\rangle_{j}$ $\leftrightarrow$ $|2\rangle_{j}$
transition with coupling constant $g$, i.e., $w_a=w_2-w_1$. $J$ is
the photon hopping strength between two coupled cavities. We obtain
the associated eigenstate and eigenenergy within zero-excitation
subspace and one-excitation subspace analytically, as shown in Table
I and Table II. The notation $|AB,CD\rangle$ represents that atom
$1$ ($2$) is in the state $|A\rangle$ ($|B\rangle$) and there are
$C$ ($D$) photons in cavity $1$ ($2$). Due to the coherent photon
hopping between two cavities, the eigenenergies in the
one-excitation subspace are shifted. The corresponding eigenstates
are
\begin{figure}[htb]
\centerline{\includegraphics[width=0.8\columnwidth]{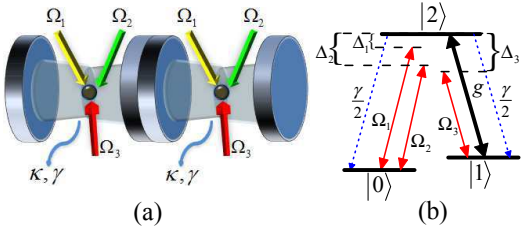}}
\caption{(Color online) (a) Experimental schematic for cooling two
identical $\Lambda$-type atoms into a maximally entangled state via
two directly coupled cavities. $\gamma$ and $\kappa$ are the atomic
spontaneous emission and the cavity decay, respectively. (b) Each
atomic level configuration with three off-resonant lasers.
$\Omega_{k}$ ($k=1,2,3$) is the $k$th laser with relevant detuning
$\Delta_{i}$.}
\end{figure}
\begin{table}\begin{center}
\caption{The eigenstates and eigenenergies of the Hamiltonian in Eq.
(1) within the zero-excitation subspace. }\label{table-1}
\begin{tabular}{lp{1in}lp{1in}}
\hline Eigenstate & Eigenenergy
\\
\hline $|\phi_{00}\rangle$ $=$ $|00,00\rangle$ & $0$
\\
\hline $|T,00\rangle$ $=$
$\frac{1}{\sqrt{2}}(|01,00\rangle+|10,00\rangle)$ & $w_{1}$
\\
\hline $|S,00\rangle$ $=$
$\frac{1}{\sqrt{2}}(|01,00\rangle-|10,00\rangle)$ & $w_{1}$
\\
\hline $|\phi_{11}\rangle$ $=$ $|11,00\rangle$ & $2w_{1}$
\\
\hline
\end{tabular}
\end{center}
\end{table}
\begin{table}\begin{center}
\caption{The eigenstates and eigenenergies of the Hamiltonian in Eq.
(\ref{e1}) within the one-excitation subspace. }\label{table-2}
\begin{tabular}{lp{2.3in}lp{2.3in}}
\hline Eigenstate & Eigenenergy
\\
\hline $|\phi_{1}\rangle$, $|\phi_{2}\rangle$ & $\lambda_{1}$ $=$
$\lambda_{2}$ $=$ $w_2+\sqrt{J^2+g^2}$
\\
\hline $|\phi_{3}\rangle$, $|\phi_{4}\rangle$ & $\lambda_{3}$ $=$
$\lambda_{4}$ $=$ $w_2-\sqrt{J^2+g^2}$
\\
\hline $|\phi_{5}\rangle$, $|\phi_{6}\rangle$ & $\lambda_{5}$ $=$
$\lambda_{6}$ $=$ $w_2$
\\
\hline $|\phi_{7}\rangle$ & $\lambda_{7}$ $=$ $w_2-w_1-J$
\\
\hline $|\phi_{8}\rangle$ & $\lambda_{8}$ $=$ $w_2-w_1+J$
\\
\hline $|\phi_{9}\rangle$ & $\lambda_{9}$ $=$
$w_1+w_2-J/2+\sqrt{J^2+4g^2}/2$
\\
\hline $|\phi_{10}\rangle$ & $\lambda_{10}$ $=$
$w_1+w_2-J/2-\sqrt{J^2+4g^2}/2$
\\
\hline $|\phi_{11}\rangle$ & $\lambda_{11}$ $=$
$w_1+w_2+J/2-\sqrt{J^2+4g^2}/2$
\\
\hline $|\phi_{12}\rangle$ & $\lambda_{12}$ $=$
$w_1+w_2+J/2+\sqrt{J^2+4g^2}/2$
\\
\hline
\end{tabular}
\end{center}
\end{table}
\begin{eqnarray}\label{e2}
|\phi_{1}\rangle&=&\frac{1}{N_{a}}\big(
\frac{\sqrt{J^2+g^2}}{g}|10,10\rangle+\frac{J}{g}|10,01\rangle+|20,00\rangle
\big),\cr |\phi_{2}\rangle&=&\frac{1}{N_{a}}\big(
\frac{J}{g}|01,10\rangle+\frac{\sqrt{J^2+g^2}}{g}|01,01\rangle+|02,00\rangle
\big),\cr |\phi_{3}\rangle&=&\frac{1}{N_{a}}\big(
\frac{J}{g}|10,01\rangle-\frac{\sqrt{J^2+g^2}}{g}|10,10\rangle+|20,00\rangle
\big),\cr |\phi_{4}\rangle&=&\frac{1}{N_{a}}\big(
\frac{J}{g}|01,10\rangle-\frac{\sqrt{J^2+g^2}}{g}|01,01\rangle+|02,00\rangle
\big),\cr |\phi_{5}\rangle&=&\frac{1}{N_{b}}\big(
-\frac{g}{J}|10,01\rangle+|20,00\rangle \big),\cr
|\phi_{6}\rangle&=&\frac{1}{N_{b}}\big(
-\frac{g}{J}|01,10\rangle+|02,00\rangle \big),\cr
|\phi_{7}\rangle&=&\frac{1}{N_{c}}\big( |00,10\rangle-|00,01\rangle
\big),\cr
|\phi_{8}\rangle&=&\frac{1}{N_{c}}\big(|00,10\rangle+|00,01\rangle
\big),\cr |\phi_{9}\rangle&=&\frac{1}{N_{d}}\big[
\frac{J-\sqrt{J^2+4g^2}}{2g}\big(|11,10\rangle-|11,01\rangle\big)\cr&&-|21,00\rangle+|12,00\rangle
\big],\cr |\phi_{10}\rangle&=&\frac{1}{N_{e}}\big[
\frac{J+\sqrt{J^2+4g^2}}{2g}\big(|11,10\rangle-|11,01\rangle\big)\cr&&-|21,00\rangle+|12,00\rangle
\big],\cr |\phi_{11}\rangle&=&\frac{1}{N_{d}}\big[
\frac{J-\sqrt{J^2+4g^2}}{2g}\big(|11,10\rangle+|11,01\rangle\big)\cr&&+|21,00\rangle+|12,00\rangle
\big],\cr |\phi_{12}\rangle&=&\frac{1}{N_{e}}\big[
\frac{J+\sqrt{J^2+4g^2}}{2g}\big(|11,10\rangle+|11,01\rangle\big)\cr&&+|21,00\rangle+|12,00\rangle
\big],
\end{eqnarray}
where $N_{a}$ $=$ $\frac{\sqrt{2(J^2+g^2)}}{g}$, $N_{b}$ $=$
$\frac{\sqrt{J^2+g^2}}{J}$, $N_{c}$ $=$ $\frac{1}{\sqrt{2}}$,
$N_{d}$ $=$ $\sqrt{\frac{1}{g^2}(J^2+4g^2-J\sqrt{J^2+4g^2})}$,
$N_{e}$ $=$ $\sqrt{\frac{1}{g^2}(J^2+4g^2+J\sqrt{J^2+4g^2})}$. To
cool the atoms into the maximally entangled state $|T\rangle$, three
optical lasers are simultaneously applied to each atom. We suppose
the $|0\rangle$ $\leftrightarrow$ $|2\rangle$ transition of each
atom is driven by two lasers with Rabi frequencies $\Omega_{m}$ and
frequencies $w_{L,m}$ ($m=1,2$), while the $|1\rangle$
$\leftrightarrow$ $|2\rangle$ transition  of each atom is driven by
another laser with Rabi frequency $\Omega_{3}$ and frequencies
$w_{L,3}$. Under the rotating wave approximation, the interaction
Hamiltonian between the atoms and lasers are described as:
\begin{eqnarray}\label{e3}
H_{AL}&=&\sum_{j=1}^{2}\sum_{m=1}^{2}(\Omega_{m}e^{iw_{L,m}t}|0\rangle_{jj}\langle
2|+H.c.)\cr&&+\sum_{j=1}^{2}(\Omega_{3}e^{iw_{L,3}t}|1\rangle_{jj}\langle
2|+H.c.).
\end{eqnarray}
Under the weak excitation condition, the probability that the system
is excited to the subspaces with more than one excitation can be
neglected. The laser-atom interaction Hamiltonian in Eq. 3 can be
expanded in terms of the eigenstates in Table I and Table II:
\begin{eqnarray}\label{e4}
H_{AL}^{'}&=&e^{iH_{NL}t}H_{AL}e^{-iH_{NL}t}\cr
&=&\sum_{x=1}^{2}\Omega_{x}\bigg[\frac{\sqrt{2}g}{2}L_{1}\sum_{k=1}^{4}e^{i(w_{L,x}-\lambda_{k})t}|00,00\rangle\langle\phi_{k}|\cr&&
+JL_{1}
\sum_{k=5}^{6}e^{i(w_{L,x}-\lambda_{k})t}|00,00\rangle\langle\phi_{k}|\cr&&
+L_{2}e^{i(w_{L,x}+w_{1}-\lambda_{9})t}|S,00\rangle\langle\phi_{9}|\cr&&
-L_{3}e^{i(w_{L,x}+w_{1}-\lambda_{10})t}|S,00\rangle\langle\phi_{10}|\cr&&
-L_{2}e^{i(w_{L,x}+w_{1}-\lambda_{10})t}|T,00\rangle\langle\phi_{11}|\cr&&
+L_{3}e^{i(w_{L,x}+w_{1}-\lambda_{12})t}|T,00\rangle\langle\phi_{12}|\bigg]\cr&&
+\Omega_{3}\bigg[\frac{g}{2}L_{1}\sum_{k=1}^{4}e^{i(w_{L,3}+w_{1}-\lambda_{k})t}|T,00\rangle\langle\phi_{k}|\cr&&
+\frac{J}{\sqrt{2}}L_{1}\sum_{k=5}^{6}e^{i(w_{L,3}+w_{1}-\lambda_{k})t}|T,00\rangle\langle\phi_{k}|\cr&&
+\frac{g}{2}L_{1}\sum_{k=1}^{4}(-1)^{k}e^{i(w_{L,3}+w_{1}-\lambda_{k})t}|S,00\rangle\langle\phi_{k}|\cr&&
+\frac{J}{\sqrt{2}}L_{1}\sum_{k=5}^{6}(-1)^{k}e^{i(w_{L,3}+w_{1}-\lambda_{k})t}|S,00\rangle\langle\phi_{k}|\cr&&
-L_{2}e^{i(w_{L,3}+2w_{1}-\lambda_{11})t}|11,00\rangle\langle\phi_{11}|+\cr&&
L_{3}e^{i(w_{L,3}+2w_{1}-\lambda_{12})t}|11,00\rangle\langle\phi_{12}|\bigg]+H.c.,
\end{eqnarray}
where $L_{1}$ $=$ $\frac{1}{\sqrt{J^2+g^2}}$, $L_{2}$ $=$
$\frac{2g^2N_{d}}{\sqrt{2J^2+8g^2}(J-\sqrt{J^2+4g^2})}$, $L_{3}$ $=$
$\frac{2g^2N_{e}}{\sqrt{2J^2+8g^2}(J+\sqrt{J^2+4g^2})}$.
\begin{figure}\label{fig2}
\centering
\includegraphics[width=0.8\columnwidth]{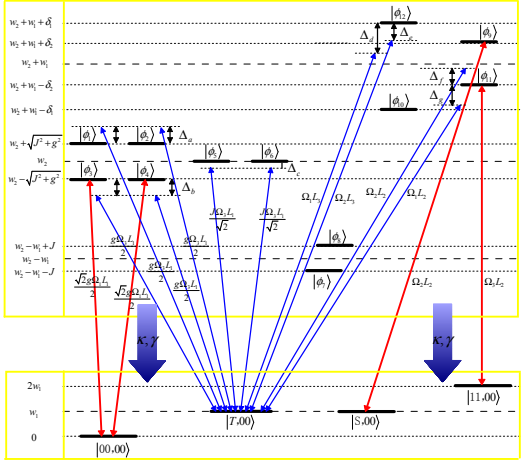} \caption{(Color
online) Level configuration in the dressed state picture, exhibiting
resonant transitions for ground states $|00,00\rangle$,
$|S,00\rangle$ and $|11,00\rangle$ by tuning three appropriate laser
frequencies, and the possible transitions induced by all the lasers
for target state $|T,00\rangle$ between the zero-excitation subspace
and one-excitation subspace. All the Rabi frequencies and detunings
of those transitions are obviously marked, where $\delta_{1}$ $=$
$\sqrt{J^2+4g^2}/2$ $+$ $J/2$, $\delta_{2}$ $=$ $\sqrt{J^2+4g^2}/2$
$-$ $J/2$, $\Delta_{a}$ $=$ $J/2$ $-$ $\sqrt{J^2+4g^2}/2$ $-$
$\sqrt{J^2+g^2}$, $\Delta_{b}$ $=$ $J/2$ $-$ $\sqrt{J^2+4g^2}/2$ $+$
$\sqrt{J^2+g^2}$, $\Delta_{c}$ $=$ $J/2$ $-$ $\sqrt{J^2+4g^2}/2$,
$\Delta_{d}$ $=$ $-$ $J/2$ $-$ $\sqrt{J^2+4g^2}/2$ $-$
$\sqrt{J^2+g^2}$, $\Delta_{e}$ $=$ $-$ $J$, $\Delta_{f}$ $=$ $-$ $J$
$+$ $\sqrt{J^2+4g^2}$, $\Delta_{g}$ $=$ $-$ $J/2$ $+$
$\sqrt{J^2+4g^2}/2$ $-$ $\sqrt{J^2+g^2}$. The solid arrows represent
one-excitation subspace dissipates into the zero-excitation subspace
induced by the cavity decay $\kappa$ and the atomic spontaneous
emission $\gamma$.}
\end{figure}
\begin{figure}\label{fig3}
\centering \subfigure[]{ \label{Fig.sub.a}
\includegraphics[width=0.8\columnwidth]{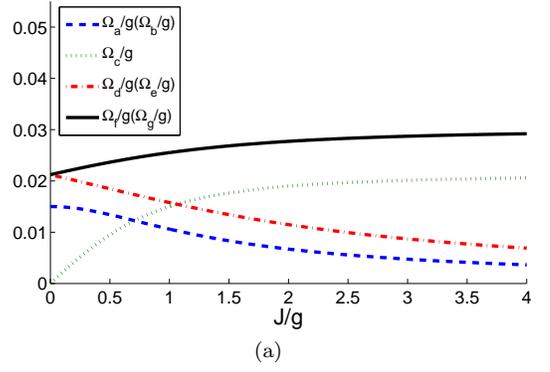}}
\subfigure[]{ \label{Fig.sub.b}
\includegraphics[width=0.8\columnwidth]{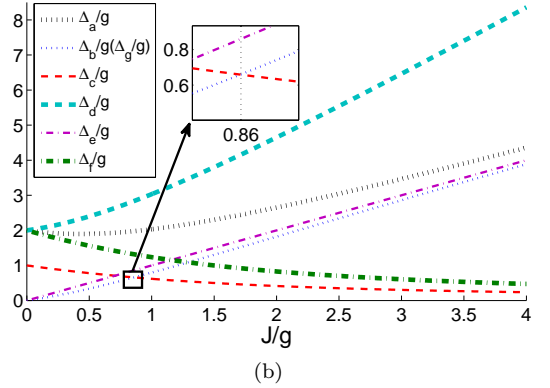}} \caption{(Color
online) The Rabi frequencies $\Omega_{x}$ $(x=a,b,c,d,e,f,g)$ with
relevant detuning $\Delta_{x}$ of the transitions for the target
state $|T,00\rangle$ is corresponding to that in Fig. 2. Different
curves versus $J/g$: (a) $\Omega_{x}/g$; (b) $\Delta_{x}/g$.}
\end{figure}
The dynamics of open dissipative system in Lindblad form is
described by the master equation
\begin{eqnarray}\label{e5}
\dot{\rho}&=&-i[(H_{NL}+H_{AL}),\rho]\cr&&
+\frac{\kappa}{2}\sum_{j=1}^{2}(2a_{j}\rho a_{j}^{\dagger}-
a_{j}^{\dagger}a_{j}\rho-\rho a_{j}^{\dagger}a_{j})\cr&&
+\frac{\gamma}{4}\sum_{j=1}^{2}\sum_{k=0}^{1}(2S_{jk}^{-}\rho
S_{jk}^{+}- S_{jk}^{+}S_{jk}^{-}\rho-\rho S_{jk}^{+}S_{jk}^{-}),
\end{eqnarray}
where $S_{jk}^{+}$ $=$ $|2\rangle_{j}\langle k|$ and $S_{jk}^{-}$
$=$ $|k\rangle_{j}\langle2|$. The master equation (5) is solved in
the subspace spanned by the eigenstates in Table I and Table II. The
fundamental physics behind the laser cooling process is the
competition between the unitary dynamics induced by the optical
lasers and the collective decays induced by the dissipation. Each
ground state can be driven to the excited states by the laser
fields, which would decay to ground states due to dissipation. The
laser frequencies are suitably chosen so that transition between the
target state $|T,00\rangle$ and each excited state of the
atom-cavity system is far off-resonant with all the applied lasers,
while each of the other ground states $|00,00\rangle$,
$|S,00\rangle$ and $|11,00\rangle$ is resonantly coupled to at least
one excited state through the laser fields. This ensures the target
state to be the unique steady state.

Assume we choose $w_{L,1}$ $=$ $w_2-\sqrt{J^2+g^2}$, $w_{L,2}$ $=$
$w_2-J/2+\sqrt{J^2+4g^2}/2$, $w_{L,3}$ $=$
$w_2-w_1+J/2-\sqrt{J^2+4g^2}/2$. When the condition $w_{1}$ $\gg$
$J+\sqrt{J^2+g^2}$, $J/2+\sqrt{J^2+4g^2}/2+\sqrt{J^2+g^2}$ is
satisfied, the detunings for the laser-driven transitions between
the target state $|T,00\rangle$ and excited states in Table II
depend on the atom-cavity coupling strength $g$ and the
cavity-cavity hopping strength $J$, as shown in Fig. 2. From the
results of Fig. 3, when $J$ is within $0.8g$ $\sim$ $1.5g$, all the
relevant detunings are much larger than the corresponding Rabi
frequencies which means the excitation of the target state
$|T,00\rangle$ is highly suppressed, while the populations of the
other ground states are quickly transferred to the excited states by
resonant laser excitations, followed by decay into the target state
$|T\rangle$ and other ground states due to dissipative dynamics.
This means that $|T,00\rangle$ is the unique steady state. Based on
the above conditions, a set of the optimal parameters is obtained by
numerically solving the full master equation (5), as shown in Fig.
4. The influences of fluctuations in $J$ and Rabi frequencies on the
fidelity are considered in Fig. 4. Even when there is $10\%$
fluctuations in $J$, the fidelity of state $|T\rangle$ only
decreases $1\%$. However, the fidelity is sensitive to the
fluctuations in Rabi frequencies. The influence of different ratios
$\gamma/\kappa$ on the fidelity is shown in Fig. 4(b), and the
optimal ratio appears near $\gamma=1.5\kappa$. For this set of
optimized experiment parameters, the populations of different ground
states versus $gt$ are plotted in Fig. 5. The numerical results show
that the target steady-state can be obtained with high fidelity even
for low cooperativity $C\sim50$, which is experimentally accessible
\cite{PRL-2004-93-233603}. This is an significant improvement as
compared to previous proposals
\cite{PRA2011-84-064302,PRL-2003-91-177901}. In conclusion, we have
proposed a feasible scheme for producing maximally entangled states
for two atoms trapped in two coupled optical cavities in the steady
state through laser cooling. Our method allows for a significant
improvement of the fidelity as compared to the previous methods for
preparation of distributed entanglement \cite{PRA2011-84-064302}.
\begin{figure}\label{fig4}
\centering
\includegraphics[width=0.9\columnwidth]{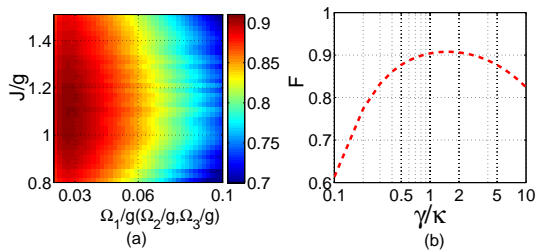} \caption{(Color
online) (a) The population of the state $|T\rangle$ versus the
photon hopping strength and the lasers strength at a finite
evolution time with $C=50$ and $\gamma=2\kappa$ . (b) The fidelity
of state $|T\rangle$ versus different ratios $\gamma/\kappa$ with
$C=50$, $w_{1}$ $=$ $8g$, $w_{2}$ $=$ $18g$, $w_{a}$ = $w_2$ $-$
$w_1$, $J$ $=$ $1.1g$, $\Omega_{1}$ $=$ $\Omega_{2}$ $=$
$\Omega_{3}$ $=$ $0.03g$ and $gt=1500$.}
\end{figure}
\begin{figure}\label{fig5}
\centering
\includegraphics[width=0.8\columnwidth]{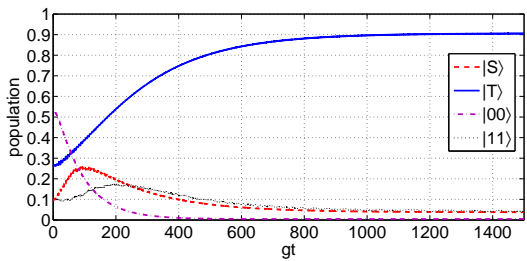} \caption{(Color
online) The populations of different ground states versus the system
evolution $gt$ with an arbitrary initial state, where $C=50$,
$\gamma=2\kappa$ and other parameters are the same as those in Fig.
4(b). }
\end{figure} L.T.S., X.Y.C., H.Z.W., and S.B.Z. acknowledge support from the
Major State Basic Research Development Program of China under Grant
No. 2012CB921601, National Natural Science Foundation of China under
Grant No. 10974028, the Doctoral Foundation of the Ministry of
Education of China under Grant No. 20093514110009, and the Natural
Science Foundation of Fujian Province under Grant No. 2009J06002.
Z.B.Y acknowledges support from the National Basic Research Program
of China under Grants No. 2011CB921200 and No. 2011CBA00200, and the
China Postdoctoral Science Foundation under Grant No. 20110490828.


\end{document}